\documentclass[%
 reprint, 
 superscriptaddress,
 amsmath,amssymb,
 aps,
 prl,
floatfix,
longbibliography
]{revtex4-2}

\usepackage{enumitem}
\usepackage{graphicx}

\usepackage{bm}
\usepackage{xcolor}
\usepackage[colorlinks=true,citecolor=blue,linkcolor=magenta]{hyperref}

\usepackage{lipsum}
\usepackage{braket}
\usepackage{quantikz}
\usepackage{rotating}

\usepackage{titlesec}
\definecolor{greybox}{HTML}{EDEDED}

\usepackage{amsthm}
\newtheorem{theorem}{Theorem}

\newtheorem{protocol}{Protocol}
\newtheorem{definition}{Definition}
\newtheorem{subroutine}{Subroutine}

\usepackage{titlesec}
\usepackage[nodisplayskipstretch]{setspace}
\definecolor{greybox}{HTML}{EDEDED}

\allowdisplaybreaks

\begin{document}

\title{Error Mitigation of BQP Computations using Measurement-Based Verification}
\author{Joseph Harris}
\email{joseph.harris@dlr.de}
\affiliation{School of Informatics, University of Edinburgh, 10 Crichton Street, Edinburgh, EH8 9AB, Scotland}
\affiliation{Institute for Software Technology, German Aerospace Center (DLR), Rathausallee 12, 53757 Sankt Augustin, Germany}
\author{Elham Kashefi}
\affiliation{School of Informatics, University of Edinburgh, 10 Crichton Street, Edinburgh, EH8 9AB, Scotland}
\affiliation{Laboratoire d’Informatique de Paris 6, CNRS, Sorbonne Université, 4 Place Jussieu, Paris 75005, France}

\date{\today}

\begin{abstract}

We present a modular error mitigation protocol for running $\mathsf{BQP}$ computations on a quantum computer with time-dependent noise. Utilising existing tools from quantum verification and measurement-based quantum computation, our protocol interleaves standard computation rounds alongside test rounds for noise sampling and inherits an exponential bound (in the number of circuit runs) on the probability that a returned classical output is correct. We introduce a post-selection technique called \textit{basketing} to address time-dependent noise and reduce overhead. The result is an error mitigation protocol which requires minimal noise assumptions, making it straightforwardly implementable on existing, NISQ devices. We perform a demonstration of the protocol using classical noisy simulation, presenting a universal measurement pattern which directly maps to (and can be tiled on) the heavy-hex layout of current IBM hardware.

\end{abstract}

\maketitle

%\tableofcontents

\section{Introduction}
\label{section:Introduction}

Quantum error mitigation (QEM) \cite{Strikis2021,Suzuki2022,Endo2018,Endo2021,Takagi2022} broadly refers to the class of techniques designed to deal with errors on near-term quantum computers where the number of qubits is too low to enable quantum error correction. Such techniques typically attempt to mitigate errors by running the same circuit or a class of modified circuits a large number of times on a noisy device and performing some classical post-analysis, for example calculating an empirical estimate of an expectation value using zero-noise extrapolation \cite{RichardsonQEMLiBenjamin, RichardsonQEMTemme, 9259940, PhysRevA.102.012426} or attempting to `undo' or `average out' errors using probabilistic error cancellation \cite{vandenBerg2023, Kim2023QEM,doi:10.1126/sciadv.aaw5686, PhysRevA.104.052607} or virtual distillation \cite{PhysRevX.11.031057,PhysRevX.11.041036, PhysRevA.107.022608, Vikstal2024studyofnoisein, PhysRevA.105.022427}. 

One factor limiting the scalability of these protocols is the necessity to adopt a noise model such that the underlying theory becomes tractable \cite{Takagi2022, takagi2022universal, Quek2024}. These noise models are typically either oversimplified and unscalable in practice or instead require an exponentially scaling classical description, all while suffering from a trade-off between quantities such as bias and variance \cite{HybridQCandQEM}. Some methods have utilised sparse noise models, but these are generally heavily-tailored to specific hardware \cite{vandenBerg2023, Kim2023QEM}. In addition, very few QEM techniques are designed to address time-dependent noise \cite{PhysRevA.106.052406}.

In this work, we present an error mitigation protocol which addresses these limitations; our protocol requires no specific noise model and instead only places basic requirements on how the noise varies with time. This enables our protocol to be straightforwardly applied to a wide range of existing NISQ devices, and for its applicability to scale with increasing future hardware complexity. Like all known error mitigation protocols, a limiting factor of our work is the exponential decay of the noiseless state -- and thus exponential increase in overhead -- with circuit depth \cite{Takagi2022, Quek2024}. Nonetheless, we demonstrate that our protocol can mitigate errors in the presence of current error rates and circuit depths.

We consider computations from the $\mathsf{BQP}$ complexity class, which can broadly be viewed as polynomially-sized decision problems with a single-bit Boolean output. A straightforward way to perform error mitigation on $\mathsf{BQP}$ computations is to run the same computation many times and to take a majority vote on the obtained outcomes. If we can bound the probability of obtaining the incorrect result below $\frac{1}{2}$, this procedure will exponentially mitigate the probability of error in the number of circuit runs. However, one cannot confidently bound this probability below $\frac{1}{2}$ in general, especially in the presence of fluctuating device noise. Moreover, since the correct result of the $\mathsf{BQP}$ computation is unknown to the user, they cannot gauge how erroneously their quantum computer is behaving. 

We show that we can often still mitigate errors in this case, with a success probability that converges exponentially to unity with the total number of device runs. To do so, we employ techniques from quantum verification, allowing us to probe the error rate of a computation despite not knowing its correct outcome. Verification refers to the situation in which we want to run a quantum computation on a remote device that we do not trust; the field has inspired error mitigation techniques before \cite{QEM-verif-postselect}. In this case we take influence from a verification protocol where runs of the standard $\mathsf{BQP}$ computation are interleaved with `test rounds' which are used for error detection \cite{verifBQP}.

We build upon this work, reducing its sensitivity to fluctuating noise by introducing a post-selection technique which we call \textit{basketing}, whereby the noise is regularly sampled and only batches of consecutive low-noise runs contribute to the final result. We show that one can use Bayesian updating to suppress the probability of failure exponentially with only a linear scaling of the total number of device shots. This technique also enables the protocol to be run in parallel across multiple QPUs/clusters at once, meaning the overhead can be spread across both time and space. 

We present our error mitigation protocol in a modular way. We first give separate subroutines for running the test and computation rounds. We also state the subroutine for a resource estimation procedure, enabling the user to specify some required target accuracy and receive optimal values of parameters for running the protocol. Finally we combine these into our error mitigation protocol, detailing how the above subroutines can be combined with our `basketing' procedure and Bayesian updating to return a classical, error-mitigated output of a $\mathsf{BQP}$ computation and an associated confidence probability. Our modular approach allows for much of the mathematical heavy-lifting to be condensed solely into the resource estimation procedure with the intention of making our work easily accessible and implementable by the wider community.

This work is presented using measurement-based quantum computation (MBQC), but can be translated to the circuit model, provided such hardware supports mid-circuit measurements in the standard basis and the ability to condition future gates on these measurement outcomes.

Finally, we demonstrate our protocol using classical simulation, presenting a universal MBQC pattern which can be mapped to and tiled on the heavy-hexagon layout of current IBM superconducting hardware. Although MBQC has a natural applicability to photonic hardware, superconducting qubits have been shown to exhibit temporal fluctuations in their stability \cite{Burnett2019, PhysRevLett.121.090502, EtxezarretaMartinez2021, martinez2022multiqubit, hirasaki2023detection, 10.1063/5.0166739} and hence our work likely has applicability to superconducting hardware too. In addition, these devices are widely accessible with competitive error rates. Under a realistic time-dependent noise model, we show that basketing enables convergence to the correct result where the underlying verification work would normally abort.

This work is structured in the following way. In Section \ref{sec:background}, we provide some necessary background theory, giving a rigorous definition of $\mathsf{BQP}$ computations and the MBQC model. In Section \ref{sec:verif-concepts} we state the subroutines used by our protocol and describe the underlying verification concepts, which are explained in further detail in the appendix. These results are repackaged from the work of \cite{verifBQP}, but we believe this a necessary step in making our work straightforward to implement. We then state the error mitigation protocol in Section \ref{sec:error-mitigation-using-verification}, describing how these subroutines can be combined with our basketing procedure and Bayesian updating to be compatible with fluctuating noise. In Section \ref{sec:numerical-example} we perform a rigorous numerical demonstration of our protocol for a 15-qubit computation. In the final sections, we conclude, discuss the limitations of our protocol and give an outlook for future work.

\section{Background theory}
\label{sec:background}
\subsection{$\mathsf{BQP}$ computations}
\label{sec:BQP}

Our protocol considers computations in the \textit{bounded-error quantum polynomial time} ($\mathsf{BQP}$) complexity class \cite{NandC}. We say a language $L$ is in $\mathsf{BQP}$ if there exists a family of polynomially-sized quantum circuits, $\{Q_n: n \in \mathbb{N} \}$  such that:
\begin{itemize}
	\item for all $n \in \mathbb{N}$, $Q_n: \{0,1\}^n \rightarrow \{0,1\}$ maps an $n$-bit input to a 1-bit output;
	\item for all $x \in L, |x| = n, \ \mathbb{P}(Q_n(x) = 1) \geq \frac{2}{3}$; 
	\item for all $x \notin L, |x| = n, \ \mathbb{P}(Q_n(x) = 0) \geq \frac{2}{3}$, 
\end{itemize}
where $\mathbb{P}(A)$ denotes the probability of an event $A$ occurring and $|x|$ denotes the number of bits in $x$. Hence, each $\mathsf{BQP}$ computation comes with an inherent error probability $p$, and consequentially each run of a $\mathsf{BQP}$ computation for a given classical input string $x$ corresponds to a \textit{non-deterministic} evaluation of $Q_{|x|}(x) = 0$ or 1, which we call the \textit{decision function}. Since $p < \frac{1}{2}$ in a noiseless scenario, we can hence estimate with exponential confidence whether $x \in L$ by evaluating $Q_{|x|}(x)$ a large number of times and taking a majority vote over the obtained outcomes. 

In this paper, we consider the implementation of a general $\mathsf{BQP}$ computation with classical input $x$ as the relevant quantum computation with quantum input $\ket{x}\ket{a} \mapsto U(\ket{x}\ket{a})$, where $\ket{a}$ is an initial ancillary state, followed by computational basis measurements to produce some classical output string $c_x$. Classical post-processing then applies some known mapping $c_x \mapsto Q_{|x|}(x)$ to infer the value of the decision function based on the measured computation outcome.

\begin{figure*}[t!]
	\hspace*{0em}
	\fboxsep4pt
	\colorbox{greybox}{\begin{minipage}{0.85\textwidth}
	\hspace*{0.5em}
	\begin{tabular}{p{0.9\linewidth}}
		\vspace*{-1em}
		\begin{subroutine}
				Computation round 
				\label{protocol:comp-round} 
				\vspace{-0.75em}
			\end{subroutine} 
			\\
		 \hline
            \vspace{0.5em}
		\textbf{Client's inputs:} A $\mathsf{BQP}$ computation with corresponding MBQC pattern $\mathcal{P}$ as defined in Section \ref{section:MBQC} and classical input $x = (x_i)_{i=1}^{|I|}$; a function $q: \{0,1\}^{|O|} \xrightarrow{} \{0,1\}$ mapping the computation outputs to the corresponding output of the decision function. \\
		\begin{spacing}{0}
			\textbf{Subroutine steps:} 
			\vspace{0.5em}
			\setlist{nolistsep}
			\begin{enumerate}[noitemsep]
				\item For each qubit with vertex $v \in V$, the Client chooses at random $\theta_v \in \Theta$ and prepares the state $\ket{+_{\theta_v}}$. The Client may also prepare $\bigotimes_{v \in I} \ket{\psi_v}$ to be the quantum input to the computation. 
				\item The Client sends the $|V|$ many prepared qubits to the Server.
                    \item The Client instructs the Server to prepare the graph state by performing a $\mathbf{CZ}_{ij}$ gate between all pairs of qubits $i,j$ with $(i,j)\in E$.
                    \item For each qubit $v \in V$, in the order dictated by the \textit{flow} $f$:
                    \begin{enumerate}
                        \item  The Client picks a uniformly random bit $r_v \in \{0,1\}$ and instructs the Server to measure the qubit at angle $\delta_v = \phi_v' + \theta_v + r_v\pi$, where $\phi_v'$ is the standard corrected angle as described in Section \ref{section:MBQC}. If $v \in I$, we replace $\theta_v$ by $\theta_v + x_v\pi$ in the above.
                        \item The Server returns the obtained `blind' measurement result $b_v$ to the Client.
                        \item The Client computes $s_v = b_v \oplus r_v$ to be the true measurement outcome. 
                    \end{enumerate}
                \item Let $\mathbf{o} = (s_i)_{i \in O}$ and set $Q_{|x|}(x) = q(\mathbf{o})$ to be the output of the decision function for this run of the $\mathsf{BQP}$ computation. 
			\end{enumerate}
		\end{spacing}
	\end{tabular}
\end{minipage}}
\caption{Subroutine for running computation rounds (adapted from \cite{verifBQP}). This procedure corresponds to running the measurement-based computation using the Universal Blind Quantum Computation framework in which the preparation and measurement angles are perturbed, analogous to a classical one-time pad.}
\label{fig:comp-round}
\end{figure*}

\begin{figure*}[p]
	\hspace*{0em}
	\fboxsep4pt
	\colorbox{greybox}{\begin{minipage}{0.85\textwidth}
	\hspace*{0.5em}
	\begin{tabular}{p{0.9\linewidth}}
		\vspace*{-1em}
		\begin{subroutine}
				Test round 
				\label{protocol:test-round} 
				\vspace{-0.75em}
			\end{subroutine} 
			\\
		 \hline
            \vspace{0.5em}
		\textbf{Client's inputs:} The graph $G$ associated with the MBQC pattern $\mathcal{P}$ for the underlying $\mathsf{BQP}$ computation; a minimal $k$-colouring $\{V_i\}_{i \in [k]}$ on $G$ for some fixed $k \in \mathbb{N}$, where $[k] = \{1, 2, \dots, k\}.$ \\
		\begin{spacing}{0}
			\textbf{Subroutine steps:} 
			\vspace{0.5em}
			\setlist{nolistsep}
			\begin{enumerate}[noitemsep]
				\item The Client picks a uniformly random integer $j \in [k]$.
                    \item For each qubit with vertex $v \in V$:
                    \begin{itemize}
                        \item If $v \in V_j$ (\textit{trap qubit}), the Client chooses a uniformly random $\theta_v \in \Theta$ and initialises $\ket{+_{\theta_v}}$.
                        \item If $v \notin V_j$ (\textit{dummy qubit}), the Client chooses a uniformly random bit $d_v \in \{0,1\}$ and prepares the computational basis state $\ket{d_v}$. 
                    \end{itemize}
                    \item The Client sends the $|V|$ many prepared qubits to the Server.
				\item The Client instructs the Server to prepare the graph  state by performing a $\mathbf{CZ}_{ij}$ gate between all pairs of qubits $i,j$ with $(i,j)\in E$.
                    \item For each qubit $v \in V$, in the order dictated by the \textit{flow} $f$, the Client instructs the Server to measure at the angle $\delta_v$ to obtain outcome $b_v$, where
                        \begin{itemize}
                            \item If $v \in V_j$, the Client chooses uniformly at random $r_v \in \{0,1\}$ and sets \newline $\delta_v = \theta_v + r_v \pi$. 
                            \item If $v \notin V_j$, the Client chooses a uniformly random angle $\delta_v \in \Theta$. 
                        \end{itemize}
                    \item Once all qubits are measured, the Client computes the Boolean expression
                    $$ \Psi \equiv \bigwedge_{v \in V_j} \left[ b_v \stackrel{\text{\tiny ?}}{=}  r_v \oplus \left( \bigoplus_{i \in N_G(v)} d_i \right) \right], $$
                    where $\bigwedge$ denotes logical AND; $A \stackrel{\text{\tiny ?}}{=} B$ denotes the Boolean expression indicating whether or not $A=B$; and $N_G(v)$ denotes the set of neighbouring vertices to $v$ in $G$. If $\Psi = \textbf{True}$, the test round returns \textbf{Pass}. If $\Psi = \textbf{False}$, the test round returns \textbf{Fail}. 
			\end{enumerate}
		\end{spacing}
	\end{tabular}
\end{minipage}}
\caption{Subroutine for running test rounds (adapted from \cite{verifBQP}). The Client chooses at random sets of trap qubits and dummy qubits from the pattern, determining the states in which the qubits are prepared. Measurement of the pattern then enables a classical post-processing check to determine whether or not the pattern was executed correctly.}
\label{fig:test-round}
\end{figure*}

\begin{figure*}[t!]
	\hspace*{0em}
	\fboxsep4pt
	\colorbox{greybox}{\begin{minipage}{0.85\textwidth}
	\hspace*{0.5em}
	\begin{tabular}{p{0.9\linewidth}}
		\vspace*{-1em}
		\begin{subroutine}
				Resource estimation 
				\label{protocol:resource-estimation} 
				\vspace{-0.75em}
			\end{subroutine} 
			\\
		 \hline
            \vspace{0.5em}
		\textbf{Client's inputs:} The Client inputs some target success probability $1-\varepsilon_\text{target}$ or some maximum total number of device runs $n$. The Client also inputs their computation as MBQC pattern $\mathcal{P}$ and the value of $k$ corresponding to the minimal possible $k$-colouring of the test round design. They also input some value of $p_\text{max}$ upper-bounding the test round error rate of the Server, obtained e.g.\ by first running a large number of consecutive test rounds. \\
		\begin{spacing}{0}
			\textbf{Subroutine steps:} 
			\vspace{0.5em}
			\setlist{nolistsep}
			\begin{enumerate}[noitemsep]
				\item If the Client inputs some target success probability $1-\varepsilon_\text{target}$:
                    \begin{enumerate}
                        \item The Client minimises the value of $n$ subject to the condition $\varepsilon_\text{max} = \varepsilon_\text{target}$ and conditions \ref{eq:eps-ver} to \ref{eq:thm-conds}, using their values of $k$ and $p$.
                        \item If the minimisation converges, the Client extracts the optimal total number of rounds $n$, proportion of test rounds $\tau$ and threshold $\Phi$. The Client sets $t = \tau n, d = (1-\tau)n$ to the nearest integers and the subroutine returns \textbf{Done}. 
                        \item If the minimisation does not converge, the subroutine returns \textbf{Abort}.
                    \end{enumerate}
                    \item If the Client inputs some target total number of device runs $n$:
                    \begin{enumerate}
                        \item The Client minimises the local-correctness upper bound $\varepsilon_\text{max}$ according to equation \ref{eq:eps-max-defn}, subject to conditions \ref{eq:eps-ver} to \ref{eq:thm-conds}, using their values of $k$ and $p$.
                        \item If the minimisation converges, the Client extracts the optimal local-correctness upper bound $\varepsilon_\text{max}$, proportion of test rounds $\tau$ and threshold $\Phi$. The Client sets $t = \tau n, d = (1-\tau)n$ to the nearest integers and the subroutine returns \textbf{Done}.
                        \item If the minimisation does not converge, the subroutine returns \textbf{Abort}.
                    \end{enumerate}
			\end{enumerate}
		\end{spacing}
	\end{tabular}
\end{minipage}}
\caption{Subroutine for running the resource estimation procedure. This corresponds to running a conditional minimisation using conditions \ref{eq:eps-ver} to \ref{eq:thm-conds} from the appendix and varies slightly depending on the Client's requirement (either to minimise the total number of device runs, or to minimise the failure probability).}
\label{fig:resource-estimation}
\end{figure*}

\subsection{Measurement-Based Quantum Computing (MBQC)}
\label{section:MBQC}

Here, we describe the key details of the MBQC framework in which our protocol is written. Readers requiring a more detailed explanation of MBQC are referred to, for example, reference \cite{MBQC}. In the appendix, we sketch a subroutine for implementing our work in the circuit model.

Measurement-based quantum computing provides an alternative universal framework to the circuit model for running quantum computations. Computations are defined using a graph $G = (V,E)$ where $V$ and $E$ are the sets of vertices and edges. The vertices correspond to qubits; we define two vertex subsets $I,O \subseteq V$ which denote the input and output qubits of the computation. We also define a list of angles $\{\phi_v\}_{v \in V}$ where the angles are quantised via $\phi_v \in \Theta \equiv \{k\pi/4\}_{k=0}^7$. The graph also has an associated \textit{flow} which dictates an ordering of the vertices -- see for example \cite{PhysRevA.74.052310} for a detailed description. We typically collate this data and call it a \textit{pattern} $\mathcal{P}$ -- see Figure \ref{fig:k-colouring} for an example.

We consider for now MBQC computations within a Client-Server scenario, where a Client (able to perform classical computations) has access to a remote Server (able to perform classical and quantum computations) via some classical channel. We assume that both the Client and Server are noise-free and that the Server is \textit{trusted} by the Client.

A computation is then performed in the following way. The Client first gives the Server the classical details of the pattern $\mathcal{P}$ corresponding to the computation. The Server then initialises the \textit{graph state},
\begin{equation}
    \ket{G} = \prod_{(i,j) \in E} \textbf{CZ}_{ij} \ \bigotimes_{v \in V} \ket{+}.
\end{equation}
This is achieved by first preparing each vertex of the graph in the $\ket{+} = \frac{1}{\sqrt2}\left( \ket{0} + \ket{1}\right)$ state, before applying a controlled-$Z$ ($\textbf{CZ}$) gate between every pair of vertices with an edge between them in the graph. For computations with a quantum input, the state $\bigotimes_{v \in I} \ket{\psi_v}$ (for arbitrary single-qubit states $\ket{\psi_v}$) may be specified by the Client to be some separate input state -- for $\mathsf{BQP}$ computations, this is typically the state $\bigotimes_{i=1}^{|x|} \ket{x_i}_i$ where $x = (x_i)_{i=1}^{|x|} \in \{0,1\}^{|x|}$ is the classical input to the computation.

Finally, the Client asks the Server to measure each qubit of $V$ in an order dictated by the flow of the graph. Each qubit with corresponding vertex $v$ is measured with respect to the basis $M_{XY}(\phi_v')$, where $M_{XY}(\theta) = \{ \ket{+_\theta}, \ket{-_\theta}\}$ with $\ket{\pm_\theta} = \frac{1}{\sqrt2}\left( \ket{0} + e^{i\theta} \ket{1}\right)$. The angles $\phi_v'$ are \textit{corrected} angles given by $\phi_v' = (-1)^{s_v^X}\phi_v + s_v^Z \pi$ where $s_v^X, s_v^Z \in \{0,1\}$ depend on the outcomes of previously measured qubits and the flow.

\subsection{Key verification concepts}
\label{sec:verif-concepts}

In this section, we describe the key verification concepts relevant to our work. These are discussed in more detail in the appendix. 

Our work utilises several concepts from an existing quantum verification protocol, in which a trusted, noiseless Client attempts to perform a computation on an untrusted, noisy Server using the measurement-based quantum computing framework \cite{verifBQP}. In the language of error mitigation, we require only a noise-independence assumption between state preparation and the rest of the computation.

We start by detailing two subroutines which our error mitigation protocol will make use of, namely the \textit{computation round} and \textit{test round} subroutines which we repackage from the work of \cite{verifBQP}. Each computation round corresponds to running our target computation on the quantum device(s) using the Universal Blind Quantum Computation protocol \cite{UBQC}. The test rounds serve the purpose of error-detection and allow us to benchmark with high accuracy the error rate of the computation rounds. We then describe the underlying verification protocol on which our work is based and state the theorem which exponentially relates the success probability to the total number of circuit runs. Finally, we give a \textit{resource estimation} subroutine which enables the Client to find optimal parameters for running our basketing error mitigation protocol. We repackage the theory in this way with the intention of making it more easily accessible and implementable by the wider community, particularly for example if the test round design is improved upon in the future.

\textit{Computation round subroutine.}
Suppose we are considering the implementation of a $\mathsf{BQP}$ computation using MBQC pattern $\mathcal{P}$ as described in Section \ref{section:MBQC}. Each computation round corresponds to running the \textit{Universal Blind Quantum Computation} protocol \cite{UBQC} on $\mathcal{P}$. We still only run one computation, but the state preparation and measurement angles are chosen by the Client such that they appear random to the Server. This \textit{blindness} is a notable strength of our work -- we perform error mitigation in a setting where the details of the computation (namely the state preparation and measurement angles) are kept completely private from the device on which it is run. The procedure required to run each computation round is given in Subroutine \ref{protocol:comp-round}. \\

\textit{Test round subroutine.} The test rounds are able to detect any general deviation from standard device behaviour for any general quantum computation, making them highly flexible for use on existing and future devices. As we will discuss later, we envision that future improvements to this work will allow the integration of more specialised test rounds which are suited to specific applications or device architectures. 

\begin{figure*}[t]
    \centering
    \includegraphics{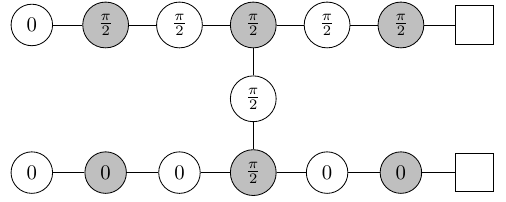}
    \caption{15-qubit MBQC pattern implementing the CNOT gate. The colour of each vertex corresponds to the $k$-colouring of the pattern with $k=2$. The angle at each vertex corresponds to the measurement angle of the associated qubit. The square vertices denote the output qubits of the computation.}
    \label{fig:k-colouring}
\end{figure*}

Suppose we are considering the implementation of a $\mathsf{BQP}$ computation using MBQC pattern $\mathcal{P}$ with graph $G = (V,E)$ as described in Section \ref{section:MBQC}. We define a \textit{$k$-colouring} $\{V_i\}_{i \in [k]}$ to be a partition of the graph $G$ into $k$ sets of vertices (called colours) such that adjacent vertices in the graph have different colours; an example is given in Figure \ref{fig:k-colouring}. With this construction, we are able to define test rounds to be run by the Server which share the same graph as $G$ but whose measurement angles dictate a computation that the Client can straightforwardly verify the correctness of using the returned classical measurement outcomes. The subroutine for running each test round is detailed in Subroutine \ref{protocol:test-round}. 

The above constructions make the test rounds and computation rounds indistinguishable to the Server due to the obfuscation property provided by the Universal Blind Quantum Computing protocol of \cite{UBQC}, which provably hides all details of the MBQC pattern (except the graph $G$, which is the same across test and computation rounds) from the Server. From an error mitigation perspective, the main benefit of this obfuscation is that these two kinds of round should share the same kinds of errors. Hence, bounding the permitted error rate of the test rounds allows us to implicitly bound the permitted error rate of the computation rounds, despite us not knowing the correct classical computation outcome. \\

We now informally describe the verification protocol that we will use and state the result which exponentially relates the success probability and the total number of circuit runs. A formal description is given in the appendix. Suppose we are given as input a $\mathsf{BQP}$ computation with a minimal $k$-colouring as described above. To run the verification protocol, we simply run $t$ test rounds and $d$ computation rounds in a uniformly random order. For each test round, we record whether it passes or fails; for each computation round with classical output $x$, we record the outcome of the decision function $Q_{|x|}(x) \in \{0,1\}$ (see Section \ref{sec:BQP}). If the proportion of failed test rounds is above some pre-determined threshold $\Phi$, the protocol aborts. If not, the protocol returns the most frequent computation result (0 or 1) from the computation rounds as the final answer alongside a probability $1 - \varepsilon$ that this result is correct. If there is no majority computation result, the protocol aborts. The quantities $\Phi$ and $\varepsilon$ are obtained using the resource estimation procedure detailed later in this section.

Now, let $p$ be the inherent error probability of the $\mathsf{BQP}$ computation as defined in Section \ref{sec:BQP}, and let $p_\text{max}$ be an upper-bound on the probability that each test round fails in the presence of noise. This can be determined theoretically using knowledge of the device noise or experimentally, for example by simply running a large number of test rounds on the device and picking $p_\text{max}$ to be larger than the proportion which fail.  With these definitions, we have the following result, stated formally as Theorem \ref{theorem:local-correctness} in the appendix.

\begin{theorem}[Exponential local-correctness (informal); adapted from Theorem 2 of \cite{verifBQP}]
\label{theorem:informal}
Assume the device noise is Markovian and round-dependent. If $p_\text{max} < \Phi < \frac{1}{k} \frac{2p-1}{2p-2}$ and the protocol returns a classical output (i.e.\ does not abort), then the probability that this output is correct is $1 - \varepsilon$ with $\varepsilon$ exponentially small in $n = t + d$, where $t$ and $d$ are the numbers of test and computation rounds. 
\end{theorem}

This theorem ensures that we can exponentially improve the success probability by increasing the total number of rounds $n$ whilst keeping $\tau = t/n$ and $1-\tau = d/n$ constant. 

The final thing we require is a method for relating the quantities $n, \tau, \Phi$ and $\varepsilon$ -- this is achieved using a resource estimation subroutine. \\

\textit{Resource estimation subroutine.} By design, we condense much of the mathematical heavy-lifting into a single resource estimation procedure which can be used to estimate the number of required device runs $n$ given some target success probability $1-\varepsilon_\text{target}$, or vice versa. In either case we run a minimiser subject to a number of parameter constraints which are detailed in the appendix. The minimiser also returns the optimal values of $\tau$ and $\Phi$. This procedure is described in Subroutine \ref{fig:resource-estimation}.

\section{Error mitigation protocol}
\label{sec:error-mitigation-using-verification}

In this section we describe our error mitigation protocol, utilising the subroutines of Section \ref{sec:verif-concepts}. This can be applied to any device with a Markovian, round-dependent noise model where the state preparation and computation noises can be deemed independent. 

Our protocol is novel but straightforward. Initially, we run a large number of (test and computation) rounds as we would for a single run of the original verification protocol. We then post-select on those periods of time where the test round pass rate is sufficiently high (and the timescale sufficiently long) to enable convergence of the minimisation procedure of Subroutine \ref{protocol:resource-estimation}. We call each such period a `basket'; we then treat each basket as a separate implementation of the verification protocol, each (if successful) returning an outcome and a success probability. The size of each basket is determined by the number of consecutive device shots over which the average test round success rate is above some pre-determined threshold. We then use a Bayesian updating procedure to combine these probabilities and infer the correct result with some known confidence that scales in the best case exponentially in the number of successful baskets. In theory, the success probability can hence be made arbitrarily close to unity by simply repeating this procedure to generate more baskets, provided we are always capable of producing baskets which return a success probability $1 - \epsilon_\text{max} > 0.5$. We state procedure this formally in Protocol \ref{protocol:error-mitigation} overleaf. \\

% \begin{center}
%      \textbf{Protocol 1: error mitigation protocol}
% \end{center}

\fboxsep4pt
	\colorbox{greybox}{\begin{minipage}{0.95\textwidth}
    \fontsize{9.5pt}{10pt}\selectfont
	\hspace*{0.5em}
	\begin{tabular}{p{0.95\linewidth}}
		\vspace*{-1em}
		\begin{protocol}
				Error mitigation protocol
				\label{protocol:error-mitigation} 
				\vspace{-0.75em}
			\end{protocol} 
			\\
		 \hline
            \vspace{0.5em}

        \fontsize{10pt}{12pt}\selectfont
		\textbf{Client's inputs:} 
        \vspace{0.5em}
        
        \fontsize{9.5pt}{10pt}\selectfont
        A $\mathsf{BQP}$ computation with corresponding MBQC pattern $\mathcal{P}$ -- as according to Section \ref{section:MBQC} -- with a minimal $k$-colouring. A definition of test rounds (Subroutine \ref{protocol:test-round}) and a corresponding resource estimation procedure (Subroutine \ref{protocol:resource-estimation}). The Client should also have knowledge of the noise behaviour of the device: firstly, a maximum tolerated error rate $\tilde{p}$ such that periods of time with this error rate (or less) can be achieved, and which enables the convergence of Subroutine \ref{protocol:resource-estimation}. Secondly, a suitable noise sampling size $T$ such that the test round failure rate at each time step $i$ is sampled from those test rounds from the time period $[i-T/2, i+T/2]$. \\
		\begin{spacing}{0}

    \vspace{0.25em}
        
    \textbf{Protocol steps:} 
    \vspace{0.5em}
    \setlist{nolistsep}
    \begin{enumerate}%[noitemsep]
    \fontsize{9.5pt}{10pt}\selectfont
    \item If the Client has some target minimum success probability $1-\varepsilon_\text{target}$, they run Subroutine \ref{protocol:resource-estimation} using $p_\text{max} = \tilde{p}$ to obtain optimal $n, \tau$ and $\Phi$ values. We set $N=n$; our goal is to produce baskets of size close to $N$, since then we expect the minimisation to converge for each individual basket. If the Client has no such target, they can pick a desired basket size $N$ freely (e.g.\ based on device access, knowledge of the noise fluctuations, etc.) and run Subroutine \ref{protocol:resource-estimation} to obtain corresponding values of $\epsilon_\text{max}, \tau$ and $\Phi$. 
    
    \item The Client picks a large number $N' \gg N$ rounds (e.g.\ $N' = 100N$) and fixes a uniformly random ordering of $\tau N'$ test and $(1-\tau)N'$ computation rounds.

    \item The Client instructs the Server to run these rounds according to the random ordering, running Subroutine \ref{protocol:comp-round} for every computation round and Subroutine \ref{protocol:test-round} for every test round. To spread the computational overhead across both space and time, the Client may optionally split these rounds into $g$ `groups' $G_i, |G_i| \geq N$, satisfying $[N'] = \bigcup_{i \in [g]} G_i$ with $G_i \cap G_j = \emptyset \ \forall \ i \neq j$ and with $i < j \implies x < y \ \forall \ x \in G_i, \ y \in G_j$ -- i.e.\ the groups contain disjoint batches of consecutive rounds. The groups may in general be run on different QPUs at different times. The Server runs these distributed computations and returns the classical measurement outcomes to the Client. 

    \item For each computation round, the Client computes the outcome of the decision function (0 or 1) according to Subroutine \ref{protocol:comp-round}. For each test round, the Client computes whether or not the test round passed according to Subroutine \ref{protocol:test-round}. 

    \item For each group $G$, the Client computes $\phi_i$, the mean test round failure rate after each round $i$ , as the average test round failure rate over the rounds $[\text{max}(i - T/2, 0), \text{min}(i+T/2, N')]$.

    \item For each group $G$, the Client uses the data $\phi_i$ to identify baskets $B_j \subseteq [N']$, $j=1,\dots,m$ of consecutive rounds where $\phi_i \leq \tilde{p} \ \forall \ i \in B_j \  \forall j$ and $|B_j| \geq N/2$. This choice in minimum basket size is arbitrary, but of course the baskets should be sufficiently large to enable convergence of Subroutine \ref{protocol:resource-estimation} whilst being small enough to enable us to identify a sufficient number of them. If no such baskets exist, the protocol returns \textbf{Abort} and the Client should restart the protocol with new parameters.

    \item For each basket $B_j$, let the values of the decision function obtained from the $d_j$ computation rounds be $\{Q^{(i)}\}_{i \in [d_j]}$; the Client takes a majority vote by computing $\mu = \sum_{i \in [d_j]} Q^{(i)}$ and setting $Q_j = 0$ if $\mu < \frac{d_j}{2}$, $Q_j = 1$ if $\mu > \frac{d_j}{2}$ or discarding the basket if $\mu = \frac{d_j}{2}$. For each of the remaining baskets $B_j$, the Client runs Subroutine \ref{protocol:resource-estimation} with $n = |B_j|$, $p_\text{max} = \tilde{p}$ and $\tau = \tau_j$ where $\tau_j$ is the proportion of test rounds in the basket. Since we expect $|B_j| \sim N$ and $\tau_j \approx \tau$, we expect the minimisation to converge. From this we obtain an associated correctness probability $1 - \varepsilon_j > 1/2$ for the outcome $Q_j$, from which we infer probabilities that 0 and 1 are the correct outcome, $q_0^{(j)}, q_1^{(j)} = \varepsilon_j, 1-\varepsilon_j$ in some order. We also obtain a maximum permitted noise threshold $\Phi_j$ and check that $f_j < \Phi_j$, where $f_j$ is the proportion of failed test rounds in basket $B_j$ -- if not, we discard the basket. Let $m' \leq m$ be the number of remaining baskets after this procedure and rename them $B_1, \dots, B_{m'}$.

    \item The Client applies Bayesian updating. Let $p_{i,j}$ denote the probability of outcome $i$ being correct after $j$ updates. Firstly, they set $p_{0,0} = p_{1,0} = 1/2$. For each remaining basket $j$, we update the associated probabilities via
    $$p_{i,j} = \frac{q_i^{(j)} p_{i,j-1}}{q_i^{(j)} p_{i,j-1} + (1-q_i^{(j)})(1-p_{i,j-1})}.$$
    \begin{itemize}
        \item If the Client has some target maximum error $\varepsilon_\text{target}$, they continue this process either until some $p_{i,j} \geq 1- \varepsilon_\text{target}$, in which case the protocol returns \textbf{True} if $i=1$ and \textbf{False} if $i = 0$ alongside the most recently updated success probability $p_{i,j}$, or until all baskets have been used. In the latter case, the Client repeats steps 2-7 and continues the Bayesian updating in step 8. If the Client has reached some maximal permitted amount of device time used without the protocol returning an outcome, the protocol returns \textbf{Abort}.
        \item If instead the Client has no target maximum error, they apply the Bayesian updating procedure until all baskets have been used. Then, let $i$ be the outcome for which $p_{i,j}$ is largest; the protocol returns \textbf{True} if $i=1$ and \textbf{False} if $i = 0$ alongside the probability $p_{i,j}$.
    \end{itemize}  
    \end{enumerate}
		\end{spacing}
	\end{tabular}
\end{minipage}}

\clearpage

% \textbf{Client's inputs:} A $\mathsf{BQP}$ computation with corresponding MBQC pattern $\mathcal{P}$ -- as according to Section \ref{section:MBQC} -- with a minimal $k$-colouring. A definition of test rounds (Subroutine \ref{protocol:test-round}) and a corresponding resource estimation procedure (Subroutine \ref{protocol:resource-estimation}). The Client should also have knowledge of the noise behaviour of the device: firstly, a maximum tolerated error rate $\tilde{p}$ such that periods of time with this error rate (or less) can be achieved, and which enables the convergence of Subroutine \ref{protocol:resource-estimation}. Secondly, a suitable noise sampling size $T$ such that the test round failure rate at each time step $i$ is sampled from those test rounds from the time period $[i-T/2, i+T/2]$. \\

The partitioning of the rounds into groups (Step 3) allows for the total overhead to be spread across both time and space, meaning the protocol does not need to be run all at once or only on one device. However, introducing too many groups can limit the potential to identify baskets, since each basket can only originate from inside a single group. In the numerical example below, we use only one group for simplicity.

\section{Numerical example}
\label{sec:numerical-example}

\begin{figure*}[t]
    \centering
    \hspace*{-0em}
    \includegraphics[scale=0.75]{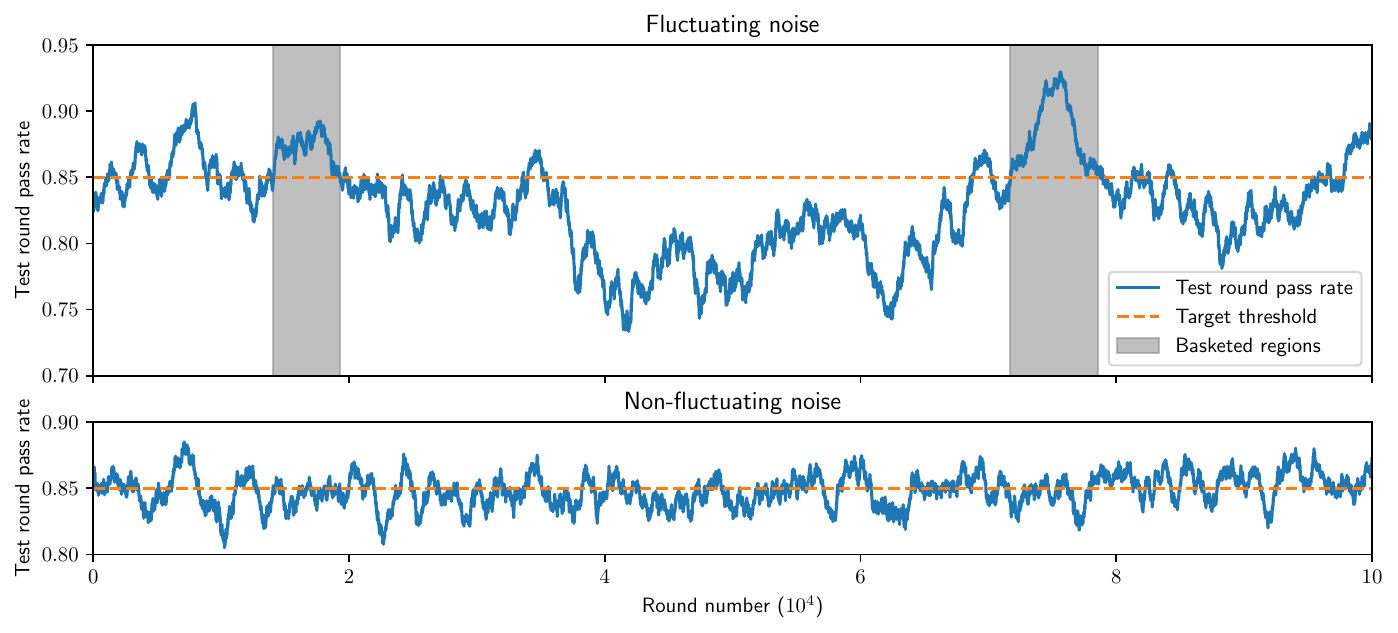}
    \caption{Plots showing the obtained average test round pass rate, sampled across the nearest 1000 rounds, for the cases of fluctuating noise (above) and constant non-fluctuating noise (below). In the former case we find two sufficiently large regions where the test round pass rate is above the threshold of 0.85; these regions become our `baskets'. In the latter case we are unable to identify any sufficiently large baskets, illustrating our protocol's suitability specifically to environments with fluctuating noise.}
    \label{fig:basketing-plot}
\end{figure*}

To demonstrate our protocol, we run noisy simulations in Qiskit \cite{qiskit2024}. The full details of these simulations are available at the associated Github repository; see Section \ref{sec:code-availability}. We note here that our use of Qiskit stems primarily from its advanced simulation capabilities, such as the ability to model mid-circuit measurements, condition the actions of future gates based on these measurement outcomes, customise and scale realistic hardware noise models, and so on. This additionally allowed us to test our protocol against realistic superconducting hardware noise; this is useful since superconducting hardware suffers from temporally fluctuating error rates \cite{Burnett2019, PhysRevLett.121.090502, EtxezarretaMartinez2021, martinez2022multiqubit, hirasaki2023detection, 10.1063/5.0166739}. The simulation of MBQC patterns using the circuit model is briefly discussed in the appendix. We note, however, that measurement-based computations are best-suited to other hardware platforms such as photonic quantum computers \cite{abughanem2024photonicquantumcomputers}, but such hardware is not widely available nor sufficiently scaled.

We model time-dependent fluctuating noise in the following way. We start by using a classical noise model designed to mimic the (simplified) behaviour of the \texttt{ibm\textunderscore sherbrooke} device. The associated noise model and device coupling map are provided as text files in the Github repository \cite{Github}. We then write a function to scale the noise up or down by adjusting the associated error probabilities according to some scalar noise parameter $s$. We find that $s=0.9$ appears to represent the boundary between being able to produce convergence within the minimisation procedure of Subroutine \ref{protocol:resource-estimation} or not, based on the maximum error rate $p_\text{max}$ of the test rounds. Thus, we model our round-parameterised noise parameter $s_i$ via a random walk in $[0.8,1]$ where $s=0.8$ represents a high noise region and $s=1$ a low noise region. Each step of the random walk is taken every 1000 device runs; this is designed to mimic device noise that varies between jobs, where each job runs a circuit with approximately $1000$ shots, but is an arbitrary choice. We also compare this to the case of a constant non-fluctuating noise level of 0.9, the midpoint of the random walk, to demonstrate that sufficiently large baskets do not generally arise without fluctuating noise. 

Our computation of interest is the 15-qubit MBQC pattern shown in Figure \ref{fig:k-colouring} with input $\ket{11}$. We choose this pattern since it contains a graph which can be directly mapped to and tiled upon the heavy-hex layout of current IBM hardware, and can realise both single and two-qubit gates. Hence we believe this pattern to be of genuine interest, since it can be tiled to realise universal blind quantum computations on existing superconducting hardware. In our specific case, the overall function of this pattern is equivalent to a CNOT and hence straightforward to verify -- we format this computation in the $\mathsf{BQP}$ language by returning \textbf{True} if the classical outcome is 10 and \textbf{False} otherwise; i.e.\ we  define $q(\mathbf{o})$ via $q(10) = 1, q(00) = q(01) = q(11) = 0$ and test the hypothesis that $\text{CNOT}\ket{11} = \ket{10}$. 

Our target is to run the 15-qubit CNOT pattern with classical input 11 and obtain the correct output 10 with 95\% confidence, i.e.\ we set the error probability bound $\varepsilon_\text{target} = 0.05$. We first start with an expected basket size $N=10,000$ and fix initial parameters $\tau = 0.9, \psi = 0.15, \varepsilon_1 = \varepsilon_2 = 0.01$ and $\varepsilon_3 = 0.1$. To two decimal places, we find the largest value of $p_\text{max}$ which produces convergence when running Subroutine \ref{protocol:resource-estimation} is $p_\text{max} = 0.15$. The minimiser returns $\tau = 0.90$.

Next, we pick $N' = 10N = 100,000$ and thus opt to run $\tau N' = 90,000$ test rounds and $(1-\tau)N' = 10,000$ computation rounds, randomly ordered, on the noisy simulator. Once completed, we calculate the mean test round failure rate at each time step according to Step 5 of the protocol; the results are shown in Figure \ref{fig:basketing-plot}. We see we are able to identify two baskets of size less than $N/2 = 5000$ rounds: basket $B_1$ from 14079 to 19277 ($N_1 = 5198$ rounds) and basket $B_2$ from 71721 to 78539  ($N_2 = 6818$ rounds). For these baskets we have $\tau_1 = \tau_2 = 0.90$ to two decimal places. In each case, a majority of the computation rounds from each basket return the correct outcome \textbf{True}. 

We then run Subroutine \ref{protocol:resource-estimation} individually for each basket, fixing the values $n = N_i$, $p=0$, $p_\text{max} = 0.15$, $k=2$ and $\tau = 0.90$. We obtain $\varepsilon_\text{max}^{(1)} = 0.17$ and $\varepsilon_\text{max}^{(2)} = 0.08$. Finally, we apply the Bayesian updating procedure, which suppresses the failure probability to 0.02 to 2 decimal places. 

Thus, we deduce that our hypothesis $\text{CNOT}\ket{11} = \ket{10}$ is correct with 98\% certainty. 

These steps are laid out with the associated program code in a Jupyter Notebook on the Github repository \cite{Github}. In the case where we had not performed error mitigation by restricting to these two baskets, we would not have been able to place an information-theoretic probability of success on our result due to the greatly increased mean error rate which prevents the minimisation procedure of Subroutine \ref{protocol:resource-estimation} from converging.

We also repeat this procedure for the case of non-fluctuating noise with a constant noise level. In this case, we see the test round pass rate stays much closer to the threshold value and as such we are unable to define any baskets of size greater than 5000 rounds. We use this to emphasise the fact that our protocol is designed for a fluctuating noise scenario in which the device is capable of extended periods of time in which the noise level is sufficiently low to enable high-quality execution. Using test rounds to regularly sample the noise, we are able to identify and post-select on these periods of time. 

\section{Conclusion, limitations and outlook}
\label{sec:conclusion-and-outlook}

In this work, we introduced a novel error mitigation protocol designed for measurement-based computations in the presence of fluctuating noise. Utilising ideas from verification, we showed how test rounds can be used to regularly sample the noise level and thus post-select on periods of time where the device noise is sufficiently low. Combining the result of Theorem \ref{theorem:informal} with a Bayesian updating procedure, we are able to exponentially improve the probability that a given answer is correct in terms of the size of each basket and their total number. Compared to the original verification protocol, we reduce its sensitivity to noise by replacing a single run with many rounds to multiple runs with fewer rounds. 

Our protocol also has several unique qualities. Blindness ensures that the details of the computation remain private to the device on which it is run. Moreover, grouping of batches of device runs enables the overhead to be spread across both time and space, meaning the error mitigation procedure can utilise multiple processors (or partitions of the same processor) simultaneously. Finally, our protocol places no specific assumption on the form of device noise, instead imposing only that the noise is Markovian, round-dependent and that state preparation and measurement noise can be considered independent. 

More generally, we hope that this work serves to demonstrate the utility and transferability of concepts such as measurement-based quantum computing, blindness and verification when used from an error mitigation perspective. 

We also take time to discuss the limitations of our work. Like all known error mitigation protocols, ours is limited by the exponential decay of the fidelity of the ideal state, and thus $p_\text{max}$, with circuit depth, or equivalently with the size of the MBQC pattern \cite{Takagi2022, Quek2024}. This means that, to avoid an exponential overhead, our work can only be scaled to deeper circuits provided there is a significant reduction in error rates. However, our numerical simulations are demonstrative that our protocol still has use in a near to intermediate-term setting. An additional limitation is that our protocol demands the use of the measurement-based model of quantum computations. This model is best-suited to photonical hardware platforms, where it is more straightforward to generate high entanglement between large numbers of qubits \cite{abughanem2024photonicquantumcomputers}. In this work, we have instead chosen to consider the implementation of a MBQC pattern on a superconducting device -- this is partially due to the availability, size, error rates and accessibility of such platforms in the near-term, and acknowledge that our work would be better demonstrated on photonic platforms once sufficiently scaled and accessible. It is also partially for this reason that we chose to consider the MBQC pattern implementing the CNOT gate, since more meaningful computations would require numbers of qubits outside the realms of classical simulation. However, we argue that this pattern is also relevant in its own right, since the underlying graph can be directly mapped to and tiled upon a heavy-hexagon qubit layout to enable the execution of universal blind computations on the IBM superconducting hardware. In the Jupyter Notebook on the associated Github repository, we show that this pattern can compete with the error rates of current hardware, thus making it useful in demonstrating the capabilities of our protocol. 

Finally, we discuss the potential for future work. We anticipate that further research will enable the development of test rounds which are tailored to specific applications or device architectures; our modular approach intentionally makes it straightforward to implement these new test rounds within a general framework for error mitigation protocols. In addition, the underlying `trappification'-based verification work is being developed to consider general computations outside of the $\mathsf{BQP}$ class and hence in the future this error mitigation framework may be able to treat these general computations as well.

\section{Code availability}
\label{sec:code-availability}

All relevant software and data used to produce the results of our demonstration in Section \ref{sec:numerical-example}, including details of the device connectivity and the noise model used, is available on Github \cite{Github}.

\section{Acknowledgments}

We thank the members of the Edinburgh \& Paris Quantum Software Lab for many useful discussions over the course of this work. 

We acknowledge the use of IBM Quantum services for
this work. The views expressed are those of the authors,
and do not reflect the official policy or position of IBM
or the IBM Quantum team. All quantum computations were
produced and simulated using Qiskit \cite{qiskit2024}.

\clearpage
\appendix

\section{Appendix}

\subsection{Verification in more detail}

Here, we describe in more detail the theory from the verification of $\mathsf{BQP}$ computations that our work inherits. Specifically, we introduce several ideas from the existing verification protocol given in \cite{verifBQP}.

This work considers the scenario of a trusted, noiseless Client interacting with an untrusted, noisy Server using the measurement-based quantum computing framework. We also assume that the Client is able to noiselessly prepare qubits and send them to the Server via a quantum channel -- for our error mitigation purposes, these assumptions can be replaced simply with a noise-independence assumption between state preparation and computation, allowing us to view the Client and Server as a single entity. For now, we keep the formalism of the original framework intact to be able to clearly identify the assumptions made.

 The verification protocol is defined by interleaving standard \textit{computation rounds} (i.e.\ individual runs of the target $\mathsf{BQP}$ computation on our quantum Server) alongside \textit{test rounds}, which are used as an error-detection mechanism to gauge whether the Server is corrupting the computation -- either maliciously or due to the inherent noise of the underlying device. The test rounds are constructed in such a way that, from the point of view of the Server, they cannot be distinguished from the standard computation rounds -- this obfuscation is achieved by forcing all test and computation rounds to share the same MBQC pattern and applying the \textit{Universal Blind Quantum Computing} protocol \cite{UBQC} to them. This forces a malicious Server -- one which is trying to corrupt and/or learn the details of our computations -- to have a high probability of corrupting some proportion $\varphi$ of the test rounds; the Client can then detect these deviations using classical post-processing on the obtained measurement results. 

To run the protocol, some uniformly random ordering of $t$ test rounds and $d$ computation rounds is selected by the Client, who instructs the Server to run them according to that order. The test rounds are recorded by the Client to either pass or fail, based on their measurement outcomes satisfying some efficiently-computable classical condition, whilst the computation rounds return a classical outcome $x$, from which we interpret the value $Q_{|x|}(x) \in \{0,1\}$ of the decision function for the target $\mathsf{BQP}$ computation (see Section \ref{sec:BQP}).

If the proportion of corrupted test rounds is at least that of some threshold proportion $\Phi$, dictated at the beginning of the protocol, then the protocol aborts. If not, the protocol takes a majority vote on the decision function values $Q_{|x|}(x) \in \{0,1\}$ obtained from the computation rounds, which are then shown in Theorem \ref{theorem:local-correctness} to exponentially concentrate on the correct result.

We emphasise here that the strength of this error-detection mechanism is derived in essence from the use of the Universal Blind Quantum Computing protocol \cite{UBQC} that enables us to hide our computations, requiring only for each qubit of the computation to be randomised by the Client. This is highly comparable to many randomised benchmarking and Pauli twirling techniques in which a random Clifford circuit is exploited to impose some depolarisation of the error channel \cite{Harper_2017,Cai2019,PhysRevLett.126.210504,PhysRevA.97.062323,PhysRevA.99.032329}.

We see in the main text that a noticeable drawback of this verification protocol within an error mitigation setting is its sensitivity to fluctuating noise -- there are many scenarios in which Subroutine \ref{protocol:resource-estimation} does not converge. Due to the security constraints, post-selection is largely prohibited and hence minor perturbations in the noise level can be viewed as malicious and can cause the protocol to abort. In addition, too high a noise level can prevent the protocol from being able to run at all. We address these issues later using our \textit{basketing} technique. 

Finally, we state the central theoretical result which our protocol utilises, placing exponential confidence on any accepted classical result using local-correctness. We first define local-correctness using the idea of a two-party protocol, as set out in the appendix of \cite{verifBQP}.

\begin{definition}[Two-party protocol]
    An $N$-round two-party protocol $\mathcal{P}_{AB}$ between an honest Client $A$ and potentially dishonest Server $B$ is a succession of $2N$ completely positive trace preserving maps $\{\mathcal{E}_i\}_{i \in \{1,\dots,N\}}$ and $\{\mathcal{F}_i\}_{i \in \{1,\dots,N\}}$ acting on $(\mathcal{A},\mathcal{C}), (\mathcal{B},\mathcal{C})$ respectively where $\mathcal{A}$ $[\mathcal{B}]$ is the register of $A$ $[B]$ and $\mathcal{C}$ is a shared communication register between them. 
\end{definition}

With this idea, we can define local-correctness.

\begin{definition}[Local-correctness]
    A two-party protocol $\mathcal{P}_{AB}$ implementing $\mathcal{U}$ for honest participants $A$ and $B$ is $\varepsilon$-locally-correct if for all possible classical inputs $x$ for $A$ we have
    \begin{equation}
        \Delta(\text{Tr}_B \circ \mathcal{P}_{AB}(\ket{x}), \ \mathcal{U}(\ket{x})) \leq \varepsilon,
    \end{equation}
    where $\Delta(\rho,\sigma) = \frac{1}{2} || \rho - \sigma ||$ for density matrices $\rho,\sigma$ and $||\rho|| = \text{Tr}\sqrt{\rho^\dagger \rho}$.
\end{definition}
Local-correctness was originally introduced in the framework of abstract cryptography \cite{MauRen11} and refers in our case to the probability that, if the protocol accepts a final value of the decision function -- as defined in Section \ref{sec:BQP} -- then we have obtained an incorrect value. As we will now see, this value can be made exponentially small. 

For a single run of the verification protocol, let $t$ and $d$ denote the total number of test and computation rounds respectively, and let $n = t + d$. The theorem below, adapted from \cite{verifBQP}, dictates that the verification protocol discussed above provides exponential convergence to the correct classical outcome as the total number of circuit runs $n$ increases.

\begin{theorem}[Exponential local-correctness; adapted from Theorem 2 of \cite{verifBQP}]
\label{theorem:local-correctness}
Assume a Markovian round-dependent model for the noise on the Client and Server devices. Let $p$ be the inherent error probability of the target $\mathsf{BQP}$ computation and let $p_\text{max}$ be an upper bound for the probability that each test round fails. If $p_\text{max} < \Phi < \frac{1}{k} \frac{2p-1}{2p-2}$, then the protocol is $\varepsilon$-locally-correct with $\varepsilon$ exponentially small in $n = t + d$. 
\end{theorem}

Here, $p$ is the inherent probability that the $\mathsf{BQP}$ computation fails in a noise-free environment (see Section \ref{sec:BQP}). Meanwhile, $p_\text{max}$ is a pre-determined upper-bound on the probability of a test round failing in the presence of device noise. This can be estimated empirically by running a large number of test rounds on the device and choosing $p_\text{max}$ to be a value larger than the mean error rate.
The manifestation of this theorem is that our computation results will exponentially concentrate on the correct value as we increase the total number of circuit runs $n$ whilst keeping the ratios $\tau \equiv \frac{t}{n}$ and $ \delta \equiv \frac{d}{n}$ fixed. \\

\textit{Details of the resource estimation subroutine.} We use the exponential local-correctness result of Theorem \ref{theorem:local-correctness} to define a resource estimation procedure in which the Client can provide either a required local-correctness upper bound $\varepsilon_\text{max}$ or some target number of device runs $n$ and receive the optimal values $t, d$ and $\Phi$ from a classical optimiser. They also have knowledge of some $p_\text{max}$, an upper-bound on the probability that a single test round fails. This can be determined theoretically using knowledge of the device noise or experimentally, for example by simply running a large number of test rounds on the device and picking $p_\text{max}$ to be larger than the proportion which fail. We now state explicitly the exponential relationship between the local-correctness $\varepsilon$ and the total number of rounds $n$. Combining details from the statements and proofs of Theorem 3, Lemma 4 and Theorem 4 from \cite{verifBQP}, we have
\begin{equation}
\varepsilon \leq \varepsilon_\text{max} \equiv  \varepsilon_\text{max}^\text{ver} + \varepsilon^\text{rej},
\label{eq:eps-max-defn}
\end{equation}
where
\begin{equation}
\resizebox{0.905\hsize}{!}{
 $\varepsilon_\text{max}^\text{ver} = \text{max} \begin{cases}
    \text{exp}\left( -2 \left( 1 - \frac{2p-1}{2p-2} + \psi - \varepsilon_3 \right) \delta \varepsilon_4^2 n \right)  \\
    \hspace{5em} + \ \text{exp} \left( - \frac{2\delta^2 \varepsilon_3^2}{\frac{2p-1}{2p-2} - \psi} n \right), \\
    \text{exp}\left( -2 \left(  \frac{2p-1}{2p-2} - \psi - \varepsilon_1 \right) \tau \varepsilon_2^2 n \right) \\
    \hspace{5em} + \ \text{exp} \left( - \frac{2\tau^2 \varepsilon_1^2}{\frac{2p-1}{2p-2} - \psi} n \right),
\end{cases} $}
\label{eq:eps-ver}\end{equation}
and
\begin{equation}
    \varepsilon^\text{rej} = \text{exp} \left( -2 (\Phi - p_\text{max})^2 \tau n \right),
    \label{eq:eps-rej}
\end{equation}
subject to the conditions
\begin{align}
    & 0 < \tau < 1, \\
    & \tau + \delta = 1, \\
    & 0 < \psi < \frac{2p-1}{2p-2}, \label{eq:c1} \\
    & 0 < \varepsilon_1 < \frac{1}{2} - \psi, \\
    & 0 < \varepsilon_2 < \frac{1}{k}, \\
    & 0 < \varepsilon_3 < \psi, \\
    & \begin{aligned} 
    & \varepsilon_4 = \left( 1 - \frac{2p-1}{2p-2} + \psi - \varepsilon_3 \right)^{-1}  \\ & \hspace{5em}  \cdot \left( \frac{1}{2} - \frac{2p-1}{2p-2} + \psi - \varepsilon_3 \right) - p, 
      \end{aligned}\\
    & \Phi = \left( \frac{1}{k} - \varepsilon_2 \right) \left( \frac{2p-1}{2p-2} - \psi - \varepsilon_1 \right),  \label{eq:Phi} \\
    & 0 \leq p_\text{max} < \Phi < \frac{1}{k} \frac{2p-1}{2p-2} \label{eq:thm-conds}.
\end{align}

Using the above relationship, we propose two possible approaches that a Client may wish to take in order to determine appropriate parameters for running our protocol and we present these approaches as a resource estimation procedure in Figure \ref{fig:resource-estimation}.

In the first, the Client chooses some preferred total number of runs $n$ and minimises the local-correctness bound $\varepsilon_\text{max}$ of equation \ref{eq:eps-max-defn} subject to the conditions \ref{eq:eps-ver} to \ref{eq:thm-conds}, with fixed input values of $n, p, p_\text{max}$ and $k$. The minimisation returns the lowest achievable accuracy $\varepsilon_\text{max}$, the corresponding maximum permitted noise threshold $\Phi$ and the value of $\tau$ dictating the proportion of test rounds to computation rounds.  

In the second and potentially more common scenario, the Client has some required accuracy $\varepsilon_\text{target}$ (in terms of a maximal permitted local-correctness value) to which they wish to perform a $\mathsf{BQP}$ computation. They instead minimise the value of $n$ subject to the condition $\varepsilon_\text{max} = \varepsilon_\text{target}$ and conditions \ref{eq:eps-ver} to \ref{eq:thm-conds}. The minimisation returns the lowest achievable value of $n$ for the Client's required accuracy, alongside the corresponding values of $\Phi$ and $\tau$ as before. This technique is used in our numerical simulations in Section \ref{sec:numerical-example}.  \\

\subsection{Simulating MBQC in the circuit model.}

Many hardware platforms do not support the direct implementation of computations defined in the measurement-based quantum computation framework. It remains possible, however, to implement our work in the circuit model provided we have the capacity for mid-circuit adaptive measurements. We emphasise that this is not the same as simply compiling the target computation in the circuit model -- for the case of 15-qubit pattern in Figure \ref{fig:k-colouring}, this would just be a single CNOT gate. This is because a direct simulation of MBQC is required to realise blindness between the test and computation rounds, a necessary step for our protocol by ensuring that the test and computation rounds share the same noise. This can be achieved by compiling each MBQC pattern $\mathcal{P}$ in the circuit model as the following sequence of gates:
\begin{itemize}
    \item A layer of $\ket{+_\theta}$ state preparations (for example using $\ket{+_\theta} = e^{i\theta Z} H\ket{0}$), where $\theta \in \left\{ \frac{k\pi}{4} \right\}_{k=0}^7 \equiv \Theta$.
    \item A layer of $CZ$ gates to initialise the graph state.
    \item For each qubit $v$ with MBQC measurement angle $\phi_v'$, apply the gate $HZ(-\phi_v')$ followed by a computational basis measurement to measure $v$ with respect to the basis $\{\ket{+_{\phi_v'}}, \ket{-_{\phi_v'}}\}$. 
\end{itemize}
This form of compilation can always be achieved since any MBQC pattern $\mathcal{P}$ can be realised in the circuit model in this form, and MBQC has been shown to be universal for quantum computation \cite{UBQC}. Indeed, we utilise this implementation in our simulation work in Section \ref{sec:numerical-example}. We leave the analysis of this compilation strategy and/or the discovery of more efficient compilation strategies to future work.

\end{document}